\documentclass[showpacs,prd,twocolumn,aps]{revtex4}
\usepackage{graphicx}
\usepackage{psfrag}
\usepackage{bm}
\usepackage{epsfig}
\usepackage{amsmath}
\usepackage{amssymb}
\usepackage{psfrag}
\begin{document}
\title{Observer dependence 
for the quantum field of phonons living on the effective curved 
space-time background of a Bose-Einstein condensate}
\title{Observer dependence 
for the phonon content of the 
sound field living on the effective curved 
space-time background of a Bose-Einstein condensate}
\author{Petr O. Fedichev$^{1,2}$ and Uwe R. Fischer$^{1,3}$}
\affiliation{$^{1}$Leopold-Franzens-Universit\"{a}t Innsbruck, 
Institut f\"{u}r Theoretische Physik, Technikerstrasse 25, 
A-6020 Innsbruck, Austria \\
$^2$Russian Research Center Kurchatov Institute, 
Kurchatov Square, 123182 Moscow, Russia \\
$^3$Eberhard-Karls-Universit\"at T\"ubingen, 
Institut f\"ur Theoretische Physik \\  
Auf der Morgenstelle 14, D-72076 T\"ubingen, Germany}

\begin{abstract} 
We demonstrate that the ambiguity of the particle
content for quantum fields
in a generally curved space-time 
can be experimentally investigated in an 
ultracold gas of atoms forming a Bose-Einstein condensate. 
We explicitly evaluate the response
of a suitable condensed matter detector, an ``Atomic Quantum Dot,''  
which can be tuned to measure time intervals associated to 
different effective acoustic 
space-times. It is found that the detector response related to 
laboratory, ``adiabatic,'' and de Sitter time intervals 
is finite in time and nonstationary, vanishing, and thermal, 
respectively.
\end{abstract}
\maketitle

\section{Introduction} 
The feat of reproducing certain features of 
the physics of relativistic 
quantum fields on curved space-time backgrounds 
\cite{unruh76,BirrellDavies},   
in the laboratory, now seems closer than ever before. Due to the 
realization of the fact that phonons or generally ``relativistic'' 
quasiparticles, propagating in a nonrelativistic background fluid,  
experience an effective curved space-time \cite{unruh,Matt,SchuetzUnruh,
PGRiemann, trautman99}, various exotic phenomena of the physics of 
classical as well as quantum fields in the curved space-times of gravity 
are getting within the 
reach of laboratory scale experiments \cite{Grisha}.

The particle content of a quantum field state in curved or flat 
space-time depends on the motional state of the observer.
A manifestation of this observer dependence is
the Unruh-Davies effect, which 
consists in the fact that a constantly accelerated 
detector in the Minkowski vacuum responds as if it were placed in a thermal 
bath with temperature proportional to its acceleration 
\cite{unruh76,BirrellDavies}. This effect has eluded observation so far:  
The value of the Unruh temperature 
$ %
T_{\rm Unruh} = 
[\hbar /(2\pi k_ {\rm B} c_{\rm L})]  a = 4\, {\rm K} \times a[10^{20} g_{\oplus}] \,,
$ 
where $a$ is the acceleration of the detector in Minkowski space 
($g_{\oplus}$ is the gravity acceleration on the surface of the Earth), and 
$c_{\rm L}$ the speed of light, makes it obvious that an observation of the 
effect is a less than trivial undertaking. Proposals for a 
 measurement with ultraintense short pulses of electromagnetic radiation have 
been put forward in, e.g., Refs. \cite{Yablonovitch,chenUnruh}.

In the following, we shall argue that the observer dependence of the particle
content of a quantum field state in curved space-time, related 
to the familiar non-uniqueness of canonical field quantization in 
Riemannian spaces \cite{fulling}, can be experimentally 
demonstrated in the readily available ultralow temperature condensed 
matter system Bose-Einstein condensate (BEC). 
More specifically, we argue that the Gibbons-Hawking 
effect \cite{Gibbons}, a curved space-time generalization of the 
Unruh-Davies effect, can be observed in an expanding BEC. To explicitly 
show that particle detection depends on the detector setting, we 
construct a condensed matter detector tuned to time intervals in 
effective laboratory and de Sitter space-times, respectively \cite{particle}. 
We show that the 
detector response is strongly different in the two cases, 
and associated to the corresponding effective space-time. 
Furthermore, we describe a system of space-time coordinates in which there is 
no particle detection whatsoever taking place, 
which we will call the ``adiabatic'' basis,
and which simply corresponds to a detector at rest in the (conformal) 
Minkowski vacuum.

\section{Hydrodynamics in an expanding cigar-shaped BEC}
It recently became apparent that 
among the most suitable systems for the simulation of quantum 
phenomena in effective space-times are Bose-Einstein condensates (BECs)
\cite{SlowLight,CSM,Leonhardt,Garay,Lidsey}. They offer the primary 
advantages of dissipation-free superflow, high controllability, with
atomic precision, of the physical parameters involved, and the accessibility of
ultralow temperatures \cite{Anglin,Picokelvin}. 
Even more importantly, in the context under consideration here, 
the theory of phononic 
quasiparticles in the inhomogeneous BEC is kinematically identical to that 
of a massless scalar field propagating on the background 
of curved space-time in $D+1$ dimensions. 

It was 
shown by Unruh that the action of the phase fluctuations $\Phi$  in a moving 
inhomogeneous superfluid may be written in the form \cite{unruh,Matt}
(we set $\hbar = m =1$, where $m$ is the mass of a superfluid 
constituent particle): 
\begin{eqnarray}
S  & = & \int d^{D+1}x 
\frac{1}{2g} \left[ -\left(\frac\partial{\partial t} \Phi -{\bm v} 
\cdot \nabla \Phi\right)^2 + c^2  (\nabla \Phi)^2 \right] \nonumber\\
& \equiv & \frac12 \int d^{D+1}x 
\sqrt{-{\sf g}} {\sf g}^{\mu\nu} 
\partial_\mu \Phi \partial_\nu \Phi \,. \label{action}
\end{eqnarray}
Here, $\bm v({\bm x},t)$ is the superfluid background velocity, 
$c({\bm x},t)=\sqrt{g\rho_0({\bm x},t)}$ is the velocity of sound, where
$g$ is a constant describing the interaction between the constituent
particles in 
the superfluid ($1/g$ is the compressibility of the fluid), 
and $\rho_0({\bm x},t)$ is the background density.
In the second line of (\ref{action}), the conventional 
hydrodynamic action is identified with the action 
of a minimally coupled scalar field in an effective curved space-time.
Furthermore, the velocity potential of the sound perturbations in the BEC 
satisfies the canonical
commutation relations of a relativistic scalar field \cite{SlowLight}.
We therefore have the exact mapping, on the level of kinematics, of 
the equation of motion for
phononic quasiparticles in a nonrelativistic superfluid,  
to quantized massless scalar fields propagating on a curved space-time  
background with local Lorentz invariance. 

\begin{center}
\begin{figure}[bt]
\psfrag{zH}{$z_{\rm H}$}
\psfrag{-zH}{$-z_{\rm H}$}
\psfrag{b}{\large $b\propto t$}
\psfrag{bperp}{\large $b_\perp\propto \sqrt t$}
\psfrag{Atomic}{Atomic}
\psfrag{Quantum Dot}{Quantum Dot}
\vspace*{1em}
\centerline{\epsfig{file=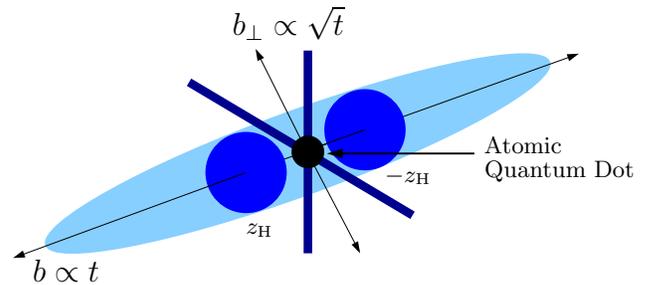,width=0.45\textwidth}}
\caption{\label{Fig1} Expansion of a cigar-shaped Bose-Einstein 
condensate.
The stationary horizon surfaces 
are located at $\pm z_{\rm H}$, respectively. 
The thick dark lines represent lasers creating an optical potential 
well in the center.} 
\end{figure}
\end{center}

To model curved space-times, 
we will consider the evolution of the BEC if we change the harmonic 
trapping frequencies with time.   
For a description of the expansion (or contraction) 
of a BEC, the so-called scaling 
solution approach is conventionally used  \cite{Scaling}. 
One starts from a cigar-shaped BEC containing 
a large number of constituent particles, i.e., which is in the 
so-called Thomas-Fermi (TF) limit \cite{BaymPethick}.  
According to \cite{Scaling}, the evolution
of a Bose-condensed atom cloud under temporal
variation of the trapping frequencies 
$\omega_\parallel(t)$ and $\omega_{\perp}(t)$ 
(in the axial and radial directions, 
respectively) can then be described by the following 
solution for the condensate wave function
\begin{equation}
\Psi = \frac{\Psi_{\rm TF}}{b_\perp {\sqrt b}} 
\exp\left[-i \int g \rho_0 ({\bm x}=0,t) dt+ i\frac{\dot b z^2}{2b}  
+ i\frac{\dot b_\perp r^2}{2b_\perp} 
\right] . \label{Wavefunction}
\end{equation}
Here, $b_{\perp}$ and $b$ are the scaling parameters describing the condensate
evolution in the radial ($\hat r$) and axial ($\hat z$) directions, 
cf.\,\,Fig.\,\ref{Fig1}.
The  initial ($b=b_\perp=1$) mean-field condensate density is given by
the usual TF expression  
\begin{equation}
|\Psi_{\rm TF}|^2=\rho_{\rm TF}(r,z)
=\rho_m \left(1-\frac{r^2}{R_{\perp}^2}-\frac{z^2}{R_\parallel^2}\right).
\end{equation}
Here, $\rho_m$ is the maximum density (in the center of the cloud) and 
the squared initial TF radii are $R^2_\parallel= 2\mu/\omega_\parallel^2 $ 
and $R^2_{\perp}=2\mu/\omega_{\perp}^2$.
The initial chemical  potential $\mu=\rho_m g$, 
where $g=4\pi a_s$, and $a_s$ is the scattering length characterizing atomic  
collisions in the (dilute) BEC. In our cylindrical 3D trap, 
we have for the initial central density 
$$
\rho_m = 
\left(\frac{6 N \omega_\perp^2 \omega_\parallel}{\sqrt 8 \pi g^{3/2}}\right)^{2/5}
.
$$
The condition that the TF condition be valid implies that 
$\mu\gg\omega_\parallel,\omega_{\perp}$. 
The solution (\ref{Wavefunction}) of the Gross-Pitaevski\v\i mean-field
equations becomes exact in this TF limit, independent of the ratio 
$\omega_\parallel/\omega_\perp$. 
However, the solution becomes exact also in the limit that 
$\omega_\perp/\omega_\parallel \rightarrow 0$, 
independent of the validity of the TF limit, the system then 
acquiring an effectively two-dimensional character 
\cite{PitaRosch,DalibardBreathing}. 
We will see below that in the opposite limit  of 
$\omega_\parallel/\omega_\perp \rightarrow 0$ there is an 
``adiabatic basis'' in which no axial excitations are created 
during the expansion, i.e., with respect to that basis there are, 
in particular, no unstable solutions possible, implying the 
stability of the expanding gas against perturbations. 
\vspace*{-1em}
\vspace*{-1.6em}

According to (\ref{Wavefunction}), the condensate density evolves as
\begin{equation}
\rho_0(r,z,t)=\frac{\rho_{\rm TF}\left({r^2}/{b_{\perp}^2},{z^2}/{b^2}\right)}
{b_{\perp}^2b}, \label{scaleddensity} 
\end{equation}
and the superfluid velocity
\begin{equation}
{\bm v}=\frac{\dot b_{\perp}}{b_{\perp}} r {\bm e}_r +\frac{\dot b}{b}z {\bm e}_z.\label{scaledv}
\end{equation}
is the gradient of the condensate  phase in 
Eq. (\ref{Wavefunction}). It increases linearly to the axial and 
radial boundaries of the condensate. 

The excitations in the limit 
$\omega_{\parallel}/\omega_\perp \rightarrow 0$  
were first studied in \cite{Zaremba}.
The description of the modes is based 
on an adiabatic separation for the axial and
longitudinal variables of the phase fluctuation field: 
\begin{equation}
\Phi (r,z,t) = \sum_n \phi_n(r) \chi_n(z,t), \label{sepansatz}
\end{equation}
where $\phi_n(r)$ is the radial wavefunction characterized by the 
quantum number $n$ (we consider only zero angular momentum modes). 
The above ansatz incorporates the fact that for strongly elongated traps the dynamics of the condensate
motion separates into a fast radial motion and a slow axial motion, which 
are essentially independent. 
The $\chi_n(z,t)$ are the mode functions for travelling wave solutions in the 
$z$ direction (plane waves for a condensate at rest read 
$\chi_n \propto \exp [- i \epsilon_{n,k} t + kz]$). The radial motion 
is assumed to be ``stiff'' such that the radial part is effectively 
time independent, because the radial time scale for adjustment of the 
density distribution after a perturbation is much less than the axial 
oscillation time scales of interest.  
The ansatz (\ref{sepansatz}) 
works independent from the ratio of healing length and 
radial size of the BEC cigar.  In the limit that the healing length is much less
than the radial size, TF wave functions are used, in the opposite limit, 
a Gaussian ansatz for the radial part of the wave function $\phi_n(r)$
is appropriate.  

For axial  excitations characterized by a wavelength $\lambda=2\pi/k$ 
exceeding the radial size $R_\perp$ of the condensate, we have $kR_\perp \ll 1$, and 
the dispersion relation reads, in the TF limit for the radial wave function
\cite{Zaremba} 
\begin{eqnarray}\label{zarembadisprel}
\epsilon^2_{n,k} & = & 
2\omega_{\perp}^2 n (n+1)+\frac{\omega_{\perp}^2}{4}(kR_{\perp})^2
\nonumber\\
& = & 2\omega_{\perp}^2 n (n+1)+c_0^2 k^2, 
\end{eqnarray}
where $c_0 =\sqrt{\mu/2} $. 
Observe that the central speed of sound $c_0$ 
is reduced by a factor $\sqrt2$ from 
the  well-established value $\sqrt{g\rho_m}=\sqrt\mu$ 
for an infinitely extended liquid \cite{Zaremba}.  

The Eqs. (\ref{sepansatz}) and (\ref{zarembadisprel}) can be 
generalized for an expanding condensate. 
Substituting in Eq.\,(\ref{sepansatz}) the rescaled radial wavefunction 
$\phi_n\equiv\phi_n(r/b_\perp)$, and inserting the result 
into the action (\ref{action}), integrating
over the radial coordinates, we find the following 
effective action for the axial modes of a given radial quantum number $n$:
\begin{eqnarray}
S_n  &= & \int dt  dz \frac{b_\perp^2 C_n(z)}{2g}
\left[ -\left(\dot \chi_n -{v}_z \partial_z\right)^2 
+ \bar c^2_n(z) (\partial_z \chi_n)^2 \right. \nonumber\\
& & \qquad  \qquad \qquad \qquad\left. + M^2_n(z)\chi_n^2 \right],
\label{1daction}
\end{eqnarray}
where the common ``conformal'' factor $C_n(z)$ is given by  
\begin{equation}
b_\perp^2 C_n(z)=\int_{r<r_m} d^2r \phi^2_n . 
\end{equation}
The integration limits are fixed by the $z$ dependent 
radial size of the cigar 
$r^2_m=R_{\perp}^2b^2_{\perp}(1-z^2/R_\parallel^2b^2)$. 
The averaged speed of sound reads 
\begin{equation}
\bar c_n^2(z,t)= \frac{g}{C_n b_\perp^2}
\int_{r<r_m} d^2r \rho_0 
\phi^2_n 
\end{equation}
and the (space and time dependent) effective mass term is, for
a given radial mode, obtained to be
\begin{equation}
M_n^2(z,t)=\frac{g}{C_n b_\perp^2} \int_{r<r_m} d^2r \rho_0 
[\partial_r\phi_n 
]^2. \label{Mass} 
\end{equation}
The phonon branch of the excitations corresponds to the gapless $n=0$ 
solution of Eq.\,(\ref{zarembadisprel}). 
In this case the  radial wavefunction $\phi_0$ does not depend on 
the radial variable $r$ \cite{Zaremba}, and the mass term vanishes, 
$M_0=0$.  
We then obtain the following expressions, 
\begin{equation}
C_0(z)=\pi R_{\perp}^2 \left(1-\frac{z^2}{b^2 R_\parallel^2}\right),
\end{equation}
and for the $z$ dependent speed of sound ($\bar c \equiv \bar c_{n=0}$):   
\begin{equation}
\bar c^2(z,t)= \frac{c^2_0}{b_{\perp}^2b}
\left(1-\frac{z^2}{b^2R_\parallel^2}\right).
\end{equation}
We will see below that we need these expressions in the limit 
$z\rightarrow 0$ only, because only in this limit we obtain 
the exact mapping of the phonon field to a quantum field
propagating in a 1+1D curved space-time.

\section{The 1+1D de Sitter metric in the condensate center}
We identify the action (\ref{1daction}) with the action of a minimally
coupled scalar field in 1+1D, according to Eq.\,(\ref{action}).
Remarkably, such an identification is possible only close to the center
$z=0$. 
The {contravariant} 1+1D metric may generally be written as 
\cite{Matt}
\begin{equation}
{\sf g}^{\mu\nu}= \frac{1}{A_c c^2}  \left( \begin{array}{cc} -1 & -v_z \\ -v_z & c^2 -v_z^2 
\end{array} \right),
\end{equation}
where $A_c$ is some arbitrary (space and time dependent)
factor and $c=\bar c (z=0)$.
Inverting this expression to get the covariant metric, we obtain
\begin{equation}
{\sf g}_{\mu\nu}= A_c \left( \begin{array}{cc} -(c^2-v_z^2) & -v_z \\ -v_z & 1
\end{array} \right). \label{gdown} 
\end{equation}
The term $\sqrt{-{\sf g}} {\sf g}^{\mu\nu}$ contained in the action
(\ref{action}) gives the familiar conformal invariance in a 1+1D space-time, 
i.e., the conformal factor $A_c$ drops out from the action and thus does 
not influence the classical equations of motion.
We therefore leave out $A_c$ in the formulae to follow, but it 
needs to be borne in mind that the metric elements 
are defined always up to the factor $A_c$. 

The actions (\ref{action}) and (\ref{1daction}) can be made consistent if 
we renormalize the phase field according to $\Phi = Z \tilde \Phi $ 
{\em and}  require that 
\begin{equation}
\frac{b_\perp^2 C_0 (0)}g Z^2 
= \frac1{\bar c} \label{ZC0}
\end{equation} 
holds.  
The factor $Z$ does not influence the equation of motion, 
but does influence the response of a detector (see section 
\ref{deSitterdetect} below).
In other words, it renormalizes the coupling of our ``relativistic field'' 
$\Phi$ to the lab frame detector. 
More explicitly, Eq.\,(\ref{ZC0}) leads to  
\begin{equation}
\frac{b_\perp}{{\sqrt b}}= 8 \sqrt{\frac{\pi}2} 
\frac1{Z^2} \sqrt{\rho_m a_s^3} 
\left(\frac{\omega_\perp}\mu\right)^2 \equiv B 
= {\rm const}. 
\label{cond1}
\end{equation} 
According to the above relation, 
we have to impose that the expansion of the cigar 
in the perpendicular direction proceeds like the square root
of the expansion in the axial direction. The constant quantity $B$ can be 
fixed externally (by the experimentalist), choosing the expansion of 
the cloud appropriately by adjusting the time dependence of the 
trapping frequencies $\omega_\parallel(t)$ and 
$\omega_\perp(t)$, according to the scaling equations
\cite{Scaling,0303063}
\begin{eqnarray}
\ddot b + \omega^2_\parallel (t) b &  = &  
\frac{\omega^2_\parallel}{b_\perp^2 b^2} = 
\frac{\omega^2_\parallel}{B^2b^{3}} , \nonumber \\ 
\ddot b_\perp + \omega^2_\perp (t) b_\perp &  = &  
\frac{\omega^2_\perp}{b_\perp^3 b} = 
\frac{\omega^2_\perp}{B^3b^{5/2}}.
\end{eqnarray}

Since both $C_0$ and $\bar c$ depend on $z$,  
an effective space-time metric for the axial phonons can be obtained 
only close to the center of the cigar-shaped condensate cloud. 
This is related to our averaging over the physical
perpendicular direction, and does not arise if the excitations are 
considered in the full D-dimensional situation, where this identification
is possible globally. Also note that the action (\ref{1daction}) 
does not contain a curvature scalar contribution 
of the form $\propto \chi_n^2 \sqrt{-{\sf g}}\, {\sf R} 
[{\sf g_{\mu\nu}}(x)]$, i.e., that it only possesses trivial
conformal invariance \cite{Dotsenko}.   

We now impose, 
in addition, the requirement that the metric is identical 
to that of a 1+1D universe, 
with a metric of the form of the de Sitter metric in 3+1D 
\cite{Gibbons,deSitter}.  
We first apply the transformation $c_0 d\tilde t = c(t) dt$ 
to the line element defined by (\ref{gdown}),  
connecting the laboratory time $t$ to the time variable $\tilde t$.
Defining $v_z/c = \sqrt \Lambda z = (B \dot b/c_0) z$ 
(note that the ``dot'' on $b$ and other quantities always refers
to ordinary laboratory time),
this results, up to the conformal factor $A_c$, in the line element  
\begin{eqnarray}
d s^2 =-c_0^2(1- \Lambda z^2)d \tilde t^2 -2c_0 z \sqrt \Lambda d\tilde t 
dz+dz^2.
\end{eqnarray}
We then apply a second transformation 
$c_0 d\tau= c_0 {d\tilde t} +z\sqrt \Lambda dz/(1-\Lambda z^2)$, with a
constant $\Lambda$. 
We are thus led to the 1+1D de Sitter metric in the form 
\cite{Gibbons}
\begin{equation}
d s^2  = -c_0^2 \left(1-\Lambda  z^2 \right)  
d {\tau}^2 + \left({1-\Lambda z^2 }\right)^{-1} 
d z^2\,. 
\label{deSitterlineelement}
\end{equation}
The transformation between $t$ and the de Sitter time $\tau$ 
(on a constant $z$ detector, such that $d\tilde t = dt $), is given by
\begin{equation}
\frac{t}{t_0} = \exp[B \dot b \tau], \label{trafo} 
\end{equation} 
where the unit of lab time $t_0 \sim \omega_\parallel^{-1}$ 
is set by the initial conditions 
for the scaling variables $b$ and $b_\perp$.   
The temperature associated with the effective 
metric (\ref{deSitterlineelement}) is the Gibbons-Hawking 
temperature \cite{Gibbons} 
\begin{eqnarray}
T_{\rm dS} 
& = &  \frac{c_0}{2\pi } \sqrt{\Lambda} = \frac B{2\pi}  {\dot b}  .
\label{TdS}
\end{eqnarray}
The ``surface gravity'' on the horizon has the value 
$a_{\rm H} = c_0^2 \sqrt \Lambda = c_0 B \dot b$, and 
the stationary horizon(s) are located at the constant values
of the $z$ coordinate 
\begin{equation}
z=\pm {z_{\rm H}}= \pm {R_\parallel} \sqrt{\frac{\omega^2_\parallel}
{2\mu\Lambda}} . \label{rH}
\end{equation}
Combining (\ref{TdS}) and (\ref{rH}), we see that $z_{\rm H}/R_\parallel$
is small if $\omega_\parallel /T_{\rm dS}\ll 4\pi $. Therefore, the  
de Sitter temperature needs to be at least of the order of
$\omega_\parallel$ for the horizon location(s) to be well inside the cloud. 
The latter condition then justifies neglecting the $z$ dependence 
in $C_0$ and $\bar c$ in Eq.\,(\ref{ZC0}).
Though there is no metric ``behind'' the horizon, i.e. at large $z$, 
this should not affect the low-energy behavior of the quantum vacuum
``outside'' the horizon.

%

\section{Determining the particle content of the quantum field}
The particle content of a quantum field state depends on the observer 
\cite{unruh76,BirrellDavies,fulling}. To detect the Gibbons-Hawking effect in de Sitter 
space, one 
has to set up a detector which measures frequencies in units of the inverse
de Sitter time $\tau$, rather than in units of the inverse laboratory time $t$.
We will show that the de Sitter time interval 
$d\tau=dt/bB=dt/{\sqrt b}b_\perp$ 
can be effectively measured by an ``Atomic Quantum Dot'' (AQD) 
\cite{AQD,0304342}.
The measured quanta can then, and only then, 
be accurately interpreted to be particles coming from a Gibbons-Hawking 
type process with a {constant} de Sitter temperature (\ref{TdS}). 
We then use the tunability to other
time intervals feasible with our detector scheme, 
and contrast this de Sitter result with what the detector ``sees'' if tuned 
to laboratory and ``adiabatic'' time intervals.

The AQD can be made in a gas of atoms possessing two hyperfine
ground states $\alpha$ and $\beta$. The atoms in the state $\alpha$ 
represent the superfluid cigar,
and are used to model the expanding universe. 
The AQD itself is formed by trapping atoms in the state $\beta$ in a 
tightly confining optical potential
$V_{\rm opt}$. The interaction of atoms in the two internal levels is
described by a set of coupling parameters $g_{cd} = 
4\pi a_{cd}$ ($c,d =\{\alpha,\beta\}$),
where $a_{cd}$ are the $s$-wave scattering
lengths characterizing short-range intra- and inter-species collisions;
$g_{\alpha\alpha}\equiv g$, $a_{\alpha\alpha} \equiv a_s$.
The on-site repulsion between the atoms $\beta$ 
in the dot is $U\sim g_{\beta \beta}/l^{3}$,
where $l$ is the characteristic size of the ground state wavefunction
of atoms $\beta$ localized in $V_{\rm opt}$. In the following, we consider the collisional
blockade limit of large $U>0$, where only one atom of type
$\beta$ can be trapped in the dot. This assumes that $U$ is larger
than all other relevant frequency scales in the dynamics of both
the AQD and the expanding superfluid. 
As a result, the collective coordinate of the AQD is modeled by a 
pseudo-spin-$1/2$, with 
spin-up/spin-down state corresponding to occupation by a single/no atom
in state $\beta$.

We first describe the AQD response to the condensate
fluctuations in the Lagrangian formalism, most familiar in a 
field theoretical context.
The detector Lagrangian takes the form 
\begin{eqnarray}
L_{\rm AQD} & = & 
i \left(\frac d{dt}{\eta^*} \right) 
\eta -\left[-\Delta+g_{\alpha\beta} (\rho_0(0,t) +\delta\rho)
\right]\eta^* \eta \nonumber \\& &
\hspace*{-2em} - \Omega \sqrt{\rho_0(0,t) l^3}
\left( 
\exp\left[-i \int_0^t g \rho_0 (0,t') dt' + i\delta\phi\right]
\eta^* \right.\label{coupling}
\nonumber \\ & & 
\left. \qquad + \exp\left[i \int_0^t g \rho_0 (0,t') dt' - i\delta\phi\right]\eta
\right).
\end{eqnarray}
Here, $\Delta$ is the detuning of the laser light 
from resonance, $ 
\rho_0(z=0,t)$ is the central mean-field part of the bath density, 
and $l$ is the size of the 
AQD ground state wave function. 
The detector variable $\eta$ is an anticommuting Grassmann variable
representing the effective spin degree of freedom of the AQD.  
The second and third lines represent the coupling of the AQD to the 
surrounding superfluid, where $\delta\phi$ and $\delta\rho$
are the fluctuating parts of the condensate phase and density 
at $z=0$, respectively.
The laser intensity and the effective transition matrix element combine
into the Rabi frequency $\Omega$; below we will make use of the fact that 
$\Omega$ can easily experimentally be changed as a function of 
laboratory time $t$, by changing the laser 
intensity with $t$. 

To simplify (\ref{coupling}), we use the canonical transformation 
\begin{eqnarray}
\eta & \rightarrow &  
\bar \eta \exp\left[-i \int_0^t g \rho_0 (0,t') dt' 
+ i\delta\phi\right]. \label{etatrafo} 
\end{eqnarray}
The above transformation amounts 
to absorbing the superfluid's chemical potential and 
the fluctuating phase  $\delta\phi$ into the wave function of the AQD, and
does not change the occupation numbers of the two AQD states.  
The transformation (\ref{etatrafo}) gives the detector Lagrangian the form 
\begin{eqnarray}
L_{\rm AQD} & = &  i 
\left(\frac{d}{dt} {{\bar \eta}^*} \right) 
\bar \eta 
- \Omega \sqrt{\rho_0(0,t) l^3}
\left(\bar \eta + \bar \eta^*\right) \label{Lbareta}  \\
& & \hspace*{-2em}
-\left[-\Delta+(g_{\alpha\beta}-g)\rho_0 (0,t)
+g_{\alpha\beta} \delta \rho 
+ \frac d{dt} \delta\phi
\right] \bar \eta^* \bar \eta\,. \nonumber
\end{eqnarray}
The laser coupling (second term in the first line)
scales as $b^{-1/2} b^{-1}_\perp$, and hence like the de Sitter time 
interval in units of the laboratory time interval, $d\tau/dt$.
We suggest to operate the detector at the time dependent detuning 
$\Delta (t) =(g_{\alpha\beta}-g)\rho_{0}(0,t)=
(g_{\alpha\beta}-g)\rho_m/(\dot b^2 B^2 t^2)$, 
which then leads to a vanishing of the first 
two terms in the square brackets of (\ref{Lbareta}).
 
We now introduce back the wave function of the AQD stemming from 
a Hamiltonian formulation, $\psi
= \psi_\beta |\beta\rangle +\psi_\alpha |\alpha\rangle $. 
An ``effective Rabi frequency'' may be defined to be 
$\omega_0=2\Omega \sqrt{\rho_m l^ 3}$;  at the detuning compensated
point, we then obtain a  
simple set of coupled equations for the AQD amplitudes
\begin{equation}
i\frac{d \psi_\beta}{d\tau}=\frac{\omega_0}2 \psi_\alpha + \delta V \psi_\beta,\,\,\,\,\qquad 
i\frac{d \psi_\alpha}{d\tau}=\frac{\omega_0}2 \psi_\beta,\label{deSitterEvol}
\end{equation}
where $\tau$ {\em is the de Sitter time.} 

We have thus shown that the detector Eqs.\,(\ref{deSitterEvol}) are  
natural evolution equations in de Sitter time $\tau$, if the Rabi 
frequency $\Omega$ is chosen to be a constant, independent of laboratory time $t$.
We will see in section \ref{detectlab} that, adjusting $\Omega$ in a certain
time dependent manner, within the same detector scheme, we can 
reproduce time intervals associated to various other effective space-times.

The coupling of the AQD to fluctuations 
in the superfluid is described by the potential
\begin{equation} 
\delta V (\tau) = \left(g_{\alpha\beta}-g \right)B b(\tau) \delta \rho(\tau).
\end{equation}
Neglecting the fluctuations in the superfluid,
the level separation implied by (\ref{deSitterEvol})
is $\omega_{0}$, and the eigenfunctions of the dressed two level
system are $|\pm\rangle= (|\alpha\rangle \pm |\beta\rangle)/\sqrt 2$. 
The quantity $\omega_0$ therefore 
plays the role of a frequency standard of the
detector. By adjusting the value of the laser intensity, one can change 
$\omega_0$, and therefore probe the response of the detector for 
various phonon frequencies. 
Note that if $g_{\alpha\beta}$ is very close to  $g$,  
to obtain the correct perturbation 
potential, higher order terms in the density 
fluctuations have to be taken into account 
in the Rabi term of (\ref{Lbareta}).

To describe the detector response, we first have to solve the 
equations of motion \,(\ref{1daction}) for the phase fluctuations, and
then evaluate the conjugate density fluctuations. 
The equation of motion  $\delta S_0 /\delta \chi_0=0$  is, 
for time independent $B$, given by
\begin{equation}
B^2 b^2 \frac d{dt} \left( b^2 \dot \chi_0 \right) - \frac1{C_0 (z_b)} 
\partial_{z_b} \left({\bar c^2(z_b) C_0 (z_b)}\partial_{z_b} 
\chi_0\right)  \label{chiEqmotion}
= 0,
\end{equation}
where $z_b = z/b$ is the scaling coordinate. Apart from the  
factor $C_0(z_b)$, stemming from averaging over the perpendicular direction,
this equation corresponds to the hydrodynamic equation of phase
fluctuations in inhomogeneous superfluids \cite{Stringari}.
At $t\rightarrow -\infty$, the condensate is in equilibrium and the
quantum vacuum phase fluctuations close to the center of the condensate can 
be written in the following form 
\begin{equation}
\hat \chi_0 =
\sqrt{\frac{g}{4C_0(0) R_\parallel \epsilon_{0,k}}}
\hat a_{k} \exp\left[
-i\epsilon_{0,k} t + ikz \right]
+ {\rm H.c.} ,
\end{equation}
where $\hat a_k,\hat a^\dagger_k$ are the annihilation and creation 
operators of a phonon. The intial quantum state of phonons is the 
ground state of the superfluid and is annihilated by the operators 
$\hat a_k$. 
With these initial conditions, the solution of (\ref{chiEqmotion}) is  
\begin{eqnarray}
\hat \chi_0 & = &
\sqrt{\frac{g}{4C_0(0) R_\parallel \epsilon_{0,k}}}
\hat a_{k} \exp\!\left[
-i\int_0^t \frac{dt' \epsilon_{0,k}}{Bb^2}+ ikz_b \right]\nonumber\\
& & + {\rm H.c.} 
\label{chi0}
\end{eqnarray}
The solution for the canonically 
conjugate density fluctuations, consequently, is 
\begin{eqnarray}
\delta\hat\rho & = & 
i \sqrt{\frac{\epsilon_{0,k}}{4C_0(0) R_\parallel g}} 
\frac{\partial}{\partial t}\! \left( 
\hat a_{k}\exp\!\left[
-i\int_0^t \frac{dt' \epsilon_{0,k}}{Bb^2}+ ikz_b \right] \right) 
\nonumber\\
& &+ {\rm H.c.} \label{deltarho} 
\end{eqnarray}
The Eqs. (\ref{chi0}) and (\ref{deltarho}) completely characterize
the $n=0$ evolution of the condensate fluctuations. Observe that the 
evolution proceeds without frequency mixing in the adiabatic time 
interval defined by $d\tau_{\rm a}= dt/Bb^2$ 
(the ``scaling time'' interval $dt/B^2b^2$ defined in \cite{0303063} 
is proportional to this adiabatic time interval).  
Therefore, in the ``adiabatic basis,'' no frequency mixing occurs and thus no 
quasiparticle excitations are created. This hints at a hidden (low energy)
symmetry,
in analogy to the (exact) 2+1D Lorentz group SO(2,1) for an isotropically
expanding BEC disk, discussed in \cite{PitaRosch}. 

\subsection{Detection in de Sitter time}  \label{deSitterdetect}
The coupling operator $\delta \hat V$ causes transitions between the 
dressed detector states $|+\rangle$ and $| -\rangle$ 
and thus can be used to effectively measure the quantum state of the phonons. 
We consider the detector response to fluctuations 
of $\hat \Psi$, by going beyond mean-field and 
using a perturbation theory in $\delta\hat V$.
There are two physically different situations. The detector 
is either at $t=0$ in its ground state,  
$(|\alpha\rangle + |\beta\rangle)/\sqrt 2$, 
or in its excited state, $(|\alpha\rangle - |\beta\rangle)/\sqrt 2$. 
We define $P_+$ and $P_-$ to be the probabilities that 
at late times $t$ the detector is excited respectively de-excited.
Using second order perturbation 
theory in $\delta\hat V$, we find that 
the transition probabilities for the detector may be written 
\begin{eqnarray}
P_{\pm} & = & \sum_k \frac{g\epsilon_{0,k}}{4R_\parallel C_0(0)} 
\left(\frac{g_{\alpha\beta}}{g}-1\right)^2 {B}^2 \left|T_\pm\right|^2 ,
\label{Ppm}
\end{eqnarray} 
where the absolute square of the transition matrix element is given by 
\begin{equation}
|T_\pm|^2 =  \left|\int_{0}^\infty \frac{d\tau}{b(\tau)} 
\exp\left[\pm i\epsilon_{0,k}\int_0^\tau 
\frac{d\tau'}{b(\tau')} + i \omega_0\tau \right]\right|^2 . \label{T(tau)}
\end{equation} 
Calculating the integrals, we obtain 
\begin{eqnarray}
P_{\pm} & = & J \left(\frac {g_{\alpha\beta}}g -1 \right)^2 {B}^2  
\frac{g\pi}{2B \dot b R_\parallel C_0(0)} 
\left\{\begin{array}{c} n_{\rm B}
\\
1+n_{\rm B} 
\end{array} ,\right.
\end{eqnarray}
where the (formally divergent) sum
\begin{equation}
J= \sum_k \frac{\omega_0}{\epsilon_{0,k}}, \label{J}
\end{equation}
and the factors 
\begin{equation}
n_{\rm B} = \frac{1}{\exp[\omega_0/T_{\rm dS}]-1}
\end{equation} 
are Bose distribution functions at the de 
Sitter temperature (\ref{TdS}).
We conclude that an expansion of the condensate in $z$ direction, with 
a constant rate faster than the harmonic 
trap oscillation frequency in that direction, 
gives an effective space-time characterized by the de Sitter temperature
$T_{\rm dS}$.

\begin{center}
\begin{figure}[t]
\psfrag{zH}{\large $z_{\rm H}$}
\psfrag{-zH}{\large $-z_{\rm H}$}
\psfrag{tau}{\large $\tau$}
\psfrag{deltaz}{\large $\delta z$}
\psfrag{z}{\large $z$}
\centerline{\epsfig{file=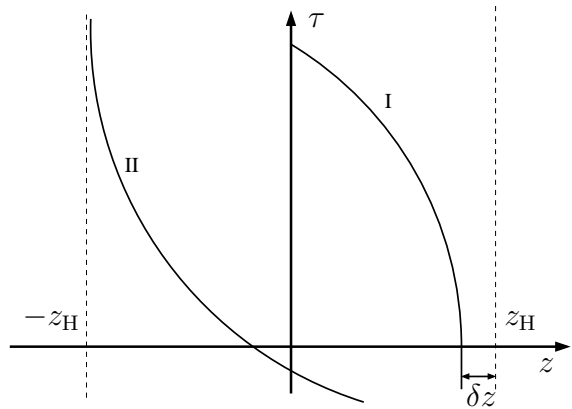,width=0.42\textwidth}}
\caption{\label{Fig3} 
Typical trajectories of phonons ($ds^2=0$)
in the de Sitter metric (\ref{deSitterlineelement}). 
The path taken by phonon I, which is at early 
times propagating near the horizon, is described  in the text. 
The path taken by phonon II, which approaches the horizon 
surface $-z_{\rm H}$ at late times $\tau$, does not lead to an 
excitation of the de Sitter detector placed at $z=0$.}
\end{figure}
\end{center}

We now show that $J$ is proportional to the total {\em de Sitter} time 
of observation, so that the probability per unit time
is a finite quantity \cite{Menshikov}. 
At late times, the detector measures phonon quanta coming, relative
to its space-time perspective, 
from close to the horizon, at a distance
$\delta z = z_{\rm H}-z \ll z_{\rm H} =\Lambda^{-1/2}$. 
The trajectory of such a phonon in the coordinates of the de Sitter metric 
(\ref{deSitterlineelement}), at late times $\tau$, is given by 
(cf. Fig.\,\ref{Fig3})
\begin{equation}
\ln \left[\frac{z_{\rm H}}{\delta z}\right] = 2\sqrt\Lambda c_0 \tau.
\end{equation}
This implies that the central AQD detector 
measures quanta which originated at the horizon with the large 
shifted frequency
\begin{equation}
\epsilon_{0,k} = \frac{\omega_0}{\sqrt 2 \Lambda^{1/4}\delta z^{1/2}} =
\frac{\omega_0}{\sqrt 2}\exp[c_0\tau \sqrt\Lambda].
\end{equation}
Making use of the above equation, we rewrite the summation over 
$k$ in (\ref{J}) as an integral over detector time:
\begin{eqnarray}
J=\sum_k \frac{\omega_0}{\epsilon_{0,k}} &  = &
\frac{R_\parallel\omega_0}{\pi c_0}\int \frac{d\epsilon_{0,k}}{\epsilon_{0,k}}
= \frac{R_\parallel\omega_0}{\pi c_0} \sqrt\Lambda c_0 \int d\tau \nonumber\\
& = &  \frac{R_\parallel\omega_0}{\pi c_0} B\dot b \tau \,.
\end{eqnarray} 
Therefore, 
the probabilities per unit detector time (de Sitter time) read, where 
upper/lower entries refer to $P_+$/$P_-$, respectively:
\begin{eqnarray}
\frac{d P_{\pm}}{d\tau} 
& = & \left(\frac{g_{\alpha\beta}}g -1 \right)^2 {B}^2   
\frac{g \omega_0}{2C_0(0) c_0} \left\{
 \begin{array}{c} n_{\rm B} 
\\
1+n_{\rm B}
\end{array} 
\right. .\label{PdtaudS}
\end{eqnarray}
They are finite quantities in the limit that $\tau\rightarrow \infty$. 
In laboratory time, the transition probabilities evolve according to 
\begin{eqnarray}
P_{\pm} (t) 
& = & P_0 
\frac{\omega_0}{T_{\rm dS}} \ln \left[\frac t{t_0} \right] 
\left\{ \begin{array}{c} n_{\rm B} 
\\
1+n_{\rm B}
\end{array} ,
\right. \label{Ppm(t)}
\end{eqnarray}
where, from relation (\ref{ZC0}), $P_0 = Z^2 [({g_{\alpha\beta}}/g-1)B]^2/2 $.
We see 
that the detector response is, as it should be, proportional to $Z^2$, 
the square of the renormalization factor of the phase fluctuation field.
\vspace*{-2.5em}

The absorption and emission coefficients ${d P_{\pm}}/{d\tau} $
satisfy Einsteinian relations. Therefore, the detector approaches
thermal equilibrium at a temperature $T_{\rm dS}$ on a time scale 
proportional to $Z^{-2} \omega_0^{-1}$. 
Our de Sitter 
AQD detector thus measures a stationary thermal 
spectrum, even though its  condensed matter 
background, with laboratory time $t$, is in a 
highly nonstationary motional state. 
Since $Z^2 \propto 
\sqrt{\rho_m a_s^3} \left({\omega_\perp}/\mu\right)^2$, 
not-too-dilute condensates with $\omega_\perp \sim \mu $ 
(i.e., close to the quasi-1D r\'egime \cite{Gorlitz})
are most suitable to observe the Gibbons-Hawking effect. 

The verification of the fact that a thermal detector 
state has been established
proceeds by the fact that the two hyperfine states $\alpha$ and $\beta$ are
spectroscopically different states of the same atom, easily  
detectable by modern quantum optical technology.
When the optical potential is switched on, the atoms are
in the empty $\alpha$ state originally, 
which is an equal-weight superposition of 
$|+\rangle = (|\alpha\rangle + |\beta\rangle)/\sqrt 2$
and $|-\rangle = (|\alpha\rangle - |\beta\rangle)/\sqrt 2$.
The thermalization due to the Gibbons-Hawking effect takes place 
in the dressed state basis consisting of the two {\em detector states}, 
i.e. of the states $|+\rangle$ and $|-\rangle$, 
on a time scale given by the quantities $P_\pm$ in Eq.\,(\ref{Ppm(t)}). 
For the laboratory observer, the Gibbons-Hawking thermal 
state will thus appear to cause damping of the Rabi oscillations 
on the thermalization timescale, i.e., friction on 
the coherent oscillating motion between the two 
detector states occurs, due to the 
thermal phonon bath perceived by the detector. 
 The occupation of the {detector states} 
can be measured directly using atomic interferometry: A $\pi/2$-pulse brings
one of them into the filled 
($\beta$) and the other into the empty ($\alpha$) state.
To increase the signal to noise ratio, one could conceive of 
manufacturing a small 
array of AQDs in a sufficiently large cigar-shaped host superfluid, 
and monitor the total population of $\beta$ atoms in this array.

\subsection{Detection in laboratory time} \label{detectlab}
We contrast the above calculation with the response 
the AQD detector would see if tuned to laboratory time. 
This can be realized if we let $\Omega \propto t$, such that
$\Omega \sqrt{\rho_0(0,t)}= \Omega \sqrt{\rho_m} / (B\dot b t) 
= {\rm const.}$, in the Rabi term on the right-hand side 
of (\ref{Lbareta}), is time 
independent. The detector has, therefore, $dt$ as its natural 
time interval in this setting. The Painlev\'e-Gullstrand metric
(\ref{gdown}) in pure laboratory frame variables, assuming
$B^2\dot b^2 \gg  \omega_\parallel^2 /b^2$
like in the derivation of the de Sitter metric 
(\ref{deSitterlineelement}), reads 
\begin{equation}
ds^2 = -\frac{c_0^2}{B^2 \dot b^2 t^2}\left(1- \Lambda z^2\right)dt^2
-\frac{2z}t dz dt +dz^2. \label{labframemetric}
\end{equation}
The metric  (\ref{labframemetric}) 
is asymptotically, for large $t$, becoming that of 
Galilei invariant ordinary 1D laboratory space, i.e. is just measuring
length along the $z$ direction, because the speed of sound
in the ever more dilute gas decreases like $1/t$ and the 
``phonon ether'' becomes increasingly less stiff.
 
The transition probabilities for absorption respectively
emission are now given by 
${\tilde P}_{\pm} =  ({g\epsilon_{0,k}}/{4R_\parallel C_0(0)}) 
\left({g_{\alpha\beta}}/{g}-1\right)^2 |\tilde T_\pm |^2$,
where the matrix elements are, cf. Eq.\,(\ref{T(tau)}), 
\begin{equation}
|\tilde T_\pm|^2 =  \left|\int_{-\infty}^\infty \frac{dt}{Bb^2} 
\exp\left[\pm i\epsilon_{0,k}\int_{-\infty}^\infty 
\frac{dt'}{Bb^2} + i \omega_0 t \right]\right|^2 .
\end{equation} 
Substituting the adiabatic time interval 
$d\tau_{\rm a}  = dt/ (B\dot b^2 t^2)$ leads for large $t$ to 
\begin{equation}
\tau_{\rm a} = \tau_{0s} -1/(B\dot b^2 t), 
\end{equation}
where $\tau_{0{\rm a}}= \int_{-\infty}^{+\infty}
dt/(B\dot b^2 t^2)$. The transformation to adiabatic time
maps $t\in [-\infty, +\infty]$ onto $\tau_{\rm a} \in 
[-\infty,\tau_{0{\rm a}}]$
and, by further substituting  
$y = \epsilon_{0,k} (\tau_{\rm a}-\tau_{0{\rm a}})$, we have 
\begin{equation}
|\tilde T_\pm|^2 = \frac{1}{\epsilon_{0k}^2} 
\left|\int_{0}^\infty dy  
\exp\left[i\left( y
\mp \frac{\omega_0 \epsilon_{0k}}{B\dot b^2} \frac 1y
\right) \right]\right|^2 .
\end{equation} 
The integral is 
a linear combination of Bessel functions.
To test its convergence properties, 
we are specifically interested in the large $\epsilon_{0k}$ 
limit. Performing a stationary phase approximation for large
$A= \pm {\omega_0 \epsilon_{0k}}/{B\dot b^2}$, we have for 
positive $A$ (absorption) that the integral above becomes
$J(A)=(\pi \sqrt A)^{1/2} \exp[-2\sqrt A]$ and 
for negative $A$ (emission) $J(A) = (\pi \sqrt{|A|})^{1/2}$. 
The final result then is 
\begin{eqnarray}
{\tilde P}_{\pm} 
&= &\left(\frac{g_{\alpha\beta}}{g}-1\right)^2  \sqrt{2\pi}  
\sqrt{\rho_m a_s^3} \left(\frac{\omega_\perp}\mu \right)^2 
\nonumber \\
& & \times \int_0^{E_{\rm Pl}}\! d\epsilon_{0k} 
\sqrt{\frac{\omega_0}{\epsilon_{0k} B \dot b^2 }} 
\times \left\{
\begin{array}{c} \exp\left[-4\sqrt{\frac{\omega_0\epsilon_{0k}}{B\dot b^2}} \,
\right]
\\ 1 
\end{array}  
\right.  \nonumber\\
&= &\left(\frac{g_{\alpha\beta}}{g}-1\right)^2  
\!\sqrt{{2\pi \rho_m a_s^3}} 
\left(\frac{\omega_\perp}\mu \right)^2 
\times \left\{
\begin{array}{c}\! \simeq \frac12 \\ 
\\\!  \sqrt{\frac{4E_{\rm Pl} \omega_0}{B \dot b^2}}
\end{array}  
\right. ,\nonumber\\
\label{labtimeP} 
\end{eqnarray}     
where $E_{\rm Pl} \sim \mu$ is the ultraviolet cutoff in the integral for 
the emission probability ${\tilde P}_-$ , the ``Planck'' scale of the 
superfluid. 
Because of the convergence of the absorption integral for ${\tilde P}_+$, the
total number of particles detected remains finite, and there are no
particles detected by the effective laboratory frame detector
at late times. This is in contrast to the de Sitter detector, which 
according to (\ref{PdtaudS}) still detects particles, in a stationary 
thermal state.


There is a detector setting which corresponds to a detector at rest
in the Minkowski vacuum.
This setting is represented by the adiabatic basis, 
with time interval defined by $d\tau_{\rm a} = dt /Bb^2$, 
realizable with the AQD by setting the Rabi frequency $\Omega \propto 1/t$. 
Then, no particles whatsoever are detected, i.e., no
frequency mixing of the positive and 
negative frequency parts of (\ref{deltarho}) 
does take place. 
The associated space-time interval 
\begin{equation} 
ds^2 = b^2[-c_0^2 d\tau_{\rm a}^2 +dz_b^2]
\end{equation}
is simply that of (conformally) flat 
Minkowski space in the spatial scaling coordinate 
$z_b$ and adiabatic time coordinate $\tau_{\rm a}$.

\section{Summary and Conclusions}

We summarize the effective space-times considered in this article, and
the associated time intervals in table \ref{table1}. 
The major observation of the present investigation is, that the  
physical nature of the effective space-time considered in the condensed matter
system reflects itself directly in the quasiparticle (phonon) content 
measured by a detector which has a natural time interval 
equal to the time interval of this particular effective space-time.
We have demonstrated that the  notion of observer dependence 
can be made experimentally 
manifest by an {\em Atomic Quantum Dot} placed at the center of a linearly 
expanding cigar-shaped 
Bose-Einstein condensate, which has a {\em tunable} effective 
time interval.  
There is thus 
opened up a possibility to 
confirm experimentally, in an effective curved space-time setting, that, 
indeed, 
``A particle detector will react to states which have 
positive frequency with respect to the detector's proper time, not with 
respect to any universal time \cite{unruh76}.''

\begin{center}
\begin{table}[t]
\begin{tabular}{|c|c|c|} \hline
\quad Effective Space \quad  & \quad Time Interval \quad 
& \quad Phonons Detected \quad \\ \hline\hline
Laboratory & $dt$ & Yes (Nonthermal) \\
de Sitter & $d\tau = dt/Bb $ & Yes (Thermal) \\
Adiabatic & $d\tau_{\rm a} = dt/Bb^2 $ & No\\
\hline
\end{tabular}
\caption{\label{table1} Various time intervals effectively 
measuring laboratory, de Sitter, and adiabatic time in a 1+1D 
BEC, respectively, 
where $b=b(t)$, and $B$ is independent of lab time.
The third entry specifies if the AQD detector, tuned 
to the given time interval, detects phonons.
In the laboratory frame, the detector has a small, nonthermal
response for a finite amount of (initial) laboratory time, 
cf. Eq.\,(\ref{labtimeP}).}
\end{table}
\end{center}
\vspace*{-1.5em}

The Gibbons-Hawking effect in the BEC 
is intrinsically quantum:  The 
signal contains a ``dimensionless Planck constant,''  
i.e. the gaseous (loop expansion) parameter $\sqrt{\rho_m a_{s}^{3}}$.
This implies that a reasonable signal-to-noise ratio can be achieved only 
by using initially dense clouds with strong interparticle interactions.
On the other hand, 
the phononic quasiparticles of the superfluid can be regarded as
non-interacting only in a first approximation in 
$\sqrt{\rho_m a_{s}^{3}}\ll 1$.
The effects of self-interaction between the phonons, induced by
larger values of the gaseous parameter, can lead to 
decoherence and the relaxation of the phonon subsystem. 
The same line of reasoning applies to the evolution
of quantum fields in the expanding universe. The interactions between
quasiparticle excitations and their connection to decoherence processes 
in cosmological models of quantum field propagation and 
particle production is, therefore, an important 
topic for future work.

\begin{acknowledgments}
We acknowledge helpful discussions with 
R. Parentani and R. Sch\"utzhold.
P.\,O.\,F. has been supported by the Austrian Science Foundation FWF and the 
Russian Foundation for Basic Research RFRR, and U.\,R.\,F. by the FWF.
They both gratefully acknowledge support from the ESF Programme  
``Cosmo\-logy in the Laboratory.''
\end{acknowledgments}

\end{document}